\documentclass[12pt,superscriptaddress,showpacs,showkeys,twocolumn,prb,aps,reprint,a4paper,floatfix]{revtex4-1}

\usepackage{graphicx}
\usepackage{amsmath, amsthm, amssymb, amsfonts}
\usepackage{float}
\usepackage{natbib}
\usepackage{physics}
\usepackage[para,online,flushleft]{threeparttable} %For tablenotes
\usepackage{array}
\usepackage{ragged2e} %For left justification

\makeatletter
\newcommand*{\citenst}[2][]{%
  \begingroup
  \let\NAT@mbox=\mbox
  \let\@cite\NAT@citenum
  \let\NAT@space\NAT@spacechar
  \let\NAT@super@kern\relax
  \renewcommand\NAT@open{[}%
  \renewcommand\NAT@close{]}%
  \citet[#1]{#2}%
  \endgroup
}
\makeatother

\makeatletter
\newcommand*{\citenumns}[2][]{%
  \begingroup
  \let\NAT@mbox=\mbox
  \let\@cite\NAT@citenum
  \let\NAT@space\NAT@spacechar
  \let\NAT@super@kern\relax
  \renewcommand\NAT@open{[}% What bracket you wish to use
  \renewcommand\NAT@close{]}%
  \cite[#1]{#2}% Here is the difference!
  \endgroup
}
\makeatother
\usepackage[caption=false]{subfig}
\begin{document}
\title{Electromagnetic models for multilayer superconducting transmission lines}
\author{Songyuan Zhao}
\email{sz311@cam.ac.uk}
\author{S. Withington}
\author{D. J. Goldie}
\author{C. N. Thomas}
\date{\today}

\affiliation{Cavendish Laboratory, JJ Thomson Avenue, Cambridge CB3 OHE, United Kingdom.}

\begin{abstract}
\noindent Thin-film superconducting transmission lines play important roles in many signal transmission and detection systems, including qubit coupling and read-out schemes, electron spin resonance systems, parametric amplifiers, and various ultra high sensitivity detectors. Here we present a rigorous method for computing the electromagnetic behaviour of superconducting microstrip transmission lines and coplanar waveguides. Our method is based on conformal mapping, and is suitable for both homogeneous superconductors and proximity-coupled multilayers. We also present an effective conductivity approximation of multilayers, thereby allowing the multilayers to be analysed using existing electromagnetic design software. We compute the numerical results for Al-Ti bilayers and discuss the validity of our full computation and homogeneous approximation.
\end{abstract}

\pacs{84.40.Az, 74.78.w, 85.25.j}
\keywords{superconducting transmission lines, conformal mapping, dispersion, microstrip, coplanar waveguide, multilayers}

\maketitle

\section{Introduction}
Thin-film transmission lines using superconducting materials have been integrated in a number of important applications to achieve low noise, high sensitivity signal transmission and detection capabilities. Sub-gap applications, focusing on frequencies below the superconductor pair-breaking threshold, benefit from high quality transmission characteristic of superconductors, while above-gap applications exploit the superconductor pair-breaking responses \citenumns{Jonas_review}. In the field of quantum computing, superconducting transmission lines are integrated in a number of proposed qubit coupling and readout schemes \citenumns{Johansson_2006, Majer_2007, Wesenberg_2009}. High quality superconducting lines have also been incorporated in electron spin resonance (ESR) systems to drive and detect spin resonance \citenumns{Pla_2016}, and to achieve quantum limited sensitivity in ESR spectroscopy \citenumns{Bienfait_2015}. In the field of detector physics, applications include travelling wave parametric amplifiers \citenumns{Eom_2012}, tunnel junction detectors \citenumns{Trucker_1985}, particle detectors in accelerators \citenumns{WEDENIG_1999}, and kinetic inductance detectors (KIDs) for astronomy observations \citenumns{Doyle_2008}, neutrino decay identifications \citenumns{CALDER_2016}, and dark matter searches \citenumns{Moore_2012}.
%Thin-film transmission lines based on superconducting materials have been integrated in a number of important applications to achieve low noise, high sensitivity signal processing capabilities. These applications include Kinetic Inductance Detectors (KIDs) \citenumns{Jonas_review} for astronomy observations and dark matter searches, travelling wave parametric amplifiers (paramps) \citenumns{Eom_2012}, SIS tunnel junction detectors \citenumns{Trucker_1985}, and qubit readout schemes \citenumns{Johansson_2006}.

Recently, there has been increased interest in proximity-coupled multilayer transmission-lines \citenumns{Catalano_2015,Songyuan_2018,Cardani_2018}. These proximity-coupled multilayers demonstrate desirable qualities such as tuneable gaps (and thus detection frequency thresholds), protection of vulnerable material through usage of self-passivating outer layers, and greater control over acoustic impedance matching \citenumns{Kaplan_1979, Cardani_2018}. Multilayers may also provide control over strain in ESR systems through combinations of materials with different thermal expansion properties \citenumns{Pla_2016,Thorbeck_2015}.

A comprehensive model of homogeneous superconducting microstrip transmission line (MTL) has previously been published by Yassin and Withington \citenumns{Withington_1995}. The model has been applied in various experimental designs and demonstrates good fit with experimental results \citenumns{Visser_2014,Yassin_2000,Shan_2007}. The model is limited to homogeneous superconducting MTL and cannot be directly applied to coplanar waveguide (CPW) devices or asymmetric multilayer devices.

In this paper, we extend the model in \citenumns{Withington_1995} to CPW geometries, as well as allowing the model to incorporate both homogeneous as well as multilayer thin-films. We also present a weighted-average approximation which allows a multilayer to be approximated by a homogeneous material with an effective conductivity, thereby allowing the inclusion of multilayers in existing design routines.

In section~\ref{sec:General}, we describe a general framework for analysing superconducting transmission lines without appealing to any particular geometry. In section~\ref{sec:Microstrip} and section~\ref{sec:CPW}, we describe the application of this framework to MTL and CPW respectively, with a detailed discussion on the conformal mapping transformations involved. We present our solution both in the form of full numerical integrals, as well as in the form of analytic approximations. Selected numerical results of the full integrals are presented in section~\ref{sec:Results}. Results demonstrate that, at sub-gap frequencies, the full multilayer solution is well approximated by the effective conductivity approximation. We summarize this work in section~\ref{sec:Conclusions}.

\section{General Analysis}
\label{sec:General}
\subsection{Surface Impedances}
\label{sec:Impedances}
In the case of homogeneous superconductors, the surface impedance $Z_{s}$ can be calculated from the BCS theory. The Mattis-Bardeen formulation \citenumns{Mattis_1958} is first used to calculate the complex conductivities $\sigma=\sigma_1-j\sigma_2$. The complex surface impedance can then be obtained using \citenumns{Kautz_1978}
\begin{align}
Z_s=\left(\frac{j\omega\mu_0}{\sigma}\right)\operatorname{coth}[(j\omega\mu_0\sigma)^{1/2}t] ,
\end{align}
where $t$ is the thickness of the homogeneous superconducting film, $\mu_0$ is the vacuum permeability, and $\omega$ is the signal angular frequency.

In the case of multilayer superconductors, the following analysis routine can be used to obtain the different surface impedances looking into the upper and lower surfaces, which are $Z_{s,u}$ and $Z_{s,l}$ respectively:
\begin{enumerate}
\item Green's functions $\theta(\hbar\omega,x)$ due to the superconducting proximity effect are found by solving the Usadel equations \citenumns{Usadel_1970} for the particular layer combinations, where $\hbar$ is the reduced Plack constant, and $x$ is the position coordinate.
\item The complex conductivities $\sigma(\hbar\omega,x)$ are found by integrating the Green's functions using  Nam's equations \citenumns{Nam_1967}.
\item Surface impedances are then found by dividing the multilayer into thin layers of thickness $\delta x$, and then cascading the resultant transfer matrices along the multilayer:
\begin{equation}
\begin{bmatrix} v_{\mathrm{s}}\\ i_{\mathrm{s}} \end{bmatrix}=\prod_{\text{all layers}} \begin{bmatrix} 1 & j\omega\mu_0 \delta x\\ \sigma(\hbar\omega,x)\delta x&1 \end{bmatrix} \begin{bmatrix} v_0\\ i_0 \end{bmatrix},
\end{equation}
where $v_0$ and $i_0$ are the potential difference and current respectively at the open boundary facing away from incoming radiation, $v_{0}/i_{0}=Z_{0}$ is the impedance of free space, $v_s$ and $i_s$ are the potential difference and current respectively at the open boundary facing towards incoming radiation, and the surface impedance is given by $v_{\mathrm{s}}/i_{\mathrm{s}}=Z_{\mathrm{s}}$. The ordering of the thin layers in the cascade determines which of $Z_{s,u}$ and $Z_{s,l}$ is obtained.
\end{enumerate}
A detailed discussion of the above methodology, as well as an analysis of numerical results for Al-Ti multilayers, can be found in \citenumns{Songyuan_2018}. We approximate the surface impedances of the multilayer side edges as $(Z_{s,u}+Z_{s,l})/2$. This first-order approximation is accurate when a multilayer approaches the homogeneous limit, where $Z_{s,u}=Z_{s,l}$. As this study focuses on thin-film superconducting transmission lines, where the edge contribution is small compared to that of upper and lower surfaces, deviation of this approximation from the full solution is small.

\begin{figure}[h]
\includegraphics[width=8.6cm]{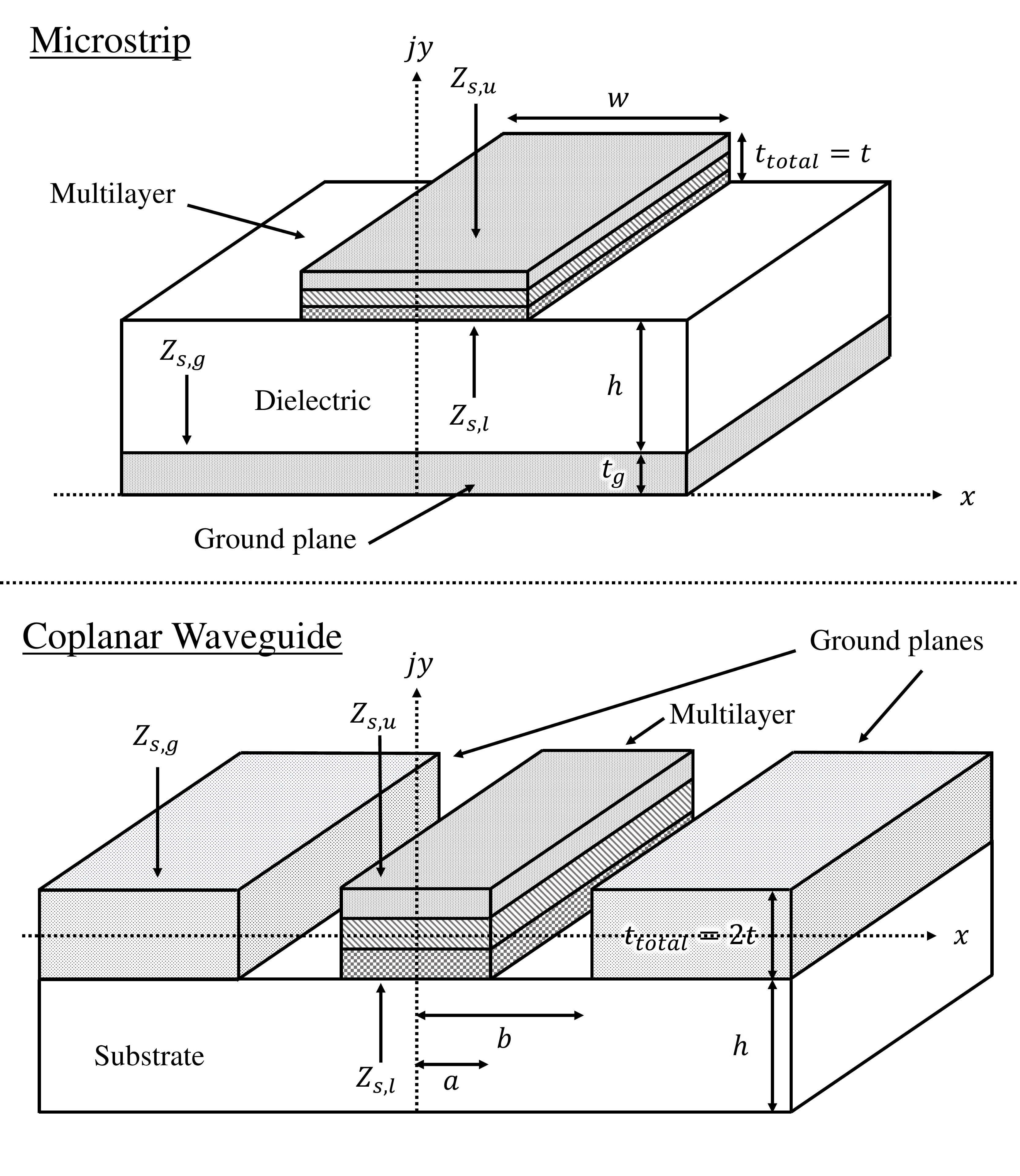}
\caption{\label{fig:Geometries} Top: geometry of a multilayer microstrip transmission line. Bottom: geometry of a multilayer coplanar waveguide.}
\end{figure}

Figure~\ref{fig:Geometries} shows geometries of multilayer MTL and CPW, as well as the surface impedances of each system. The surface impedances looking into upper and lower sides of a multilayer, in general, are not the same, $Z_{s,u} \neq Z_{s,l}$. Previous theories of superconducting transmission lines do not take into account this difference in surface impedances. Thus, extensions of existing theories have to be made to properly analyse multilayer transmission line systems.

\subsection{Transmission Line}
The general forms of the series impedance and shunt admittance of a transmission line are given by \citenumns{Whitaker_1988}
\begin{align}
\label{eqn:Z}
Z &=j(k_0\eta_0)g_1 +2\sum_{n}g_{2,n}Z_{s,n} \\
\label{eqn:Y}
Y &=j\left(\frac{k_0}{\eta_0}\right)\left(\frac{\epsilon_{fm}}{g_1}\right) \, ,
\end{align}
where $k_0$ is the free-space wavenumber, $\eta_0$ is the impedance of free-space, subscript $n$ identifies superconductor surfaces, which in later sections are upper, lower, and ground surfaces, denoted by subscripts $u,l,$ and $g$ respectively, $\epsilon_{fm}$ is the effective modal dielectric constant, which is given by existing normal conductor transmission line theories, for example \citenumns{Edwards_1976, Gupta_1996}. $g_1$ and $g_2$ are geometric factors to be determined in subsequent sections for specific geometries. The characteristic impedance is given by $\eta = (Z/Y)^{1/2}$. The propagation constant is given by $\gamma=\alpha+j\beta=({ZY})^{1/2}$, where $\alpha$ is the attenuation constant and $\beta$ is the phase constant. This allows us to calculate the overall superconducting effective dielectric constant $\epsilon_{\text{eff}}=\left({c\gamma}/{\omega}\right)^2$, and the loss tangent $\operatorname{tan}\delta=-{\operatorname{Im}(\epsilon_{\text{eff}})}/{\operatorname{Re}(\epsilon_{\text{eff}})}$.

$g_1$ is characteristic of the field distribution external to the transmission line, and is related to the quasi-static capacitance of the transmission line \citenumns{Chang_1979} by
\begin{align}
\label{eqn:g1}
g_1=\frac{\epsilon_0}{C_{\text{vac}}} \, ,
\end{align}
where $C_{\text{vac}}$ is the in vacuo capacitance when the dielectric and substrate of the transmission line have their relative permittivity values set to unity.

$g_2$ is characteristic of the field penetration into the superconductor \citenumns{Withington_1995}. We now introduce a new variable
\begin{equation}
  \psi_{n}={g_{2,n}}/{g_1} \, . \label{eq:Psi_defn}
\end{equation} We compare the attenuation constant obtained from $\alpha=\operatorname{Re}[({ZY})^{1/2}]$ against $\alpha$ obtained from integrating the electric fields \citenumns{Assadourian_1952},
\begin{align}
\alpha &= \sum_{n}\frac{P_{c,n}}{2P} \\
&=\frac{{\epsilon_{fm}}^{1/2}}{2\eta_0} \sum_{n}\frac{\int_{n}|E(z)|^2ds}{\int_{ext}|E(z)|^2dA}\,R_{s,n} \, ,
\end{align}
where $P_c$ is the power dissipated in the conductors per unit length, $P$ is the time-averaged power flow across the transmission line cross-section, $E(z)$ is the electric field in $z$-plane, the top integral of length element $ds$ is along the cross-section of superconductor surface $n$, and the bottom integral of area element $dA$ is across all area external to the superconductors. The cross-sections of a MTL and a CPW are shown in the top subfigures of figure~\ref{fig:conformalMicrostrip} and \ref{fig:conformalCPW1} respectively. From this comparison, we observe that
\begin{align}
\label{eqn:Psi}
\psi_{n} = \frac{1}{2}\frac{\int_{n}|E(z)|^2ds}{\int_{ext}|E(z)|^2dA} \,.
\end{align}

The dispersion and attenuation characteristics of a superconducting transmission line are fully known once $g_1$ and $\psi_n$ are found using appropriate conformal transformations, and $Z_s$ using the methods described in section~\ref{sec:Impedances}.

Care must be taken when performing calculations to find $\psi_n$ using conformal mapping, since the power loss integration in the numerator of equation~(\ref{eqn:Psi}) is not invariant under a conformal transformation. The following relevant results are derived in \citenumns{Assadourian_1952}:
\begin{equation}
|E(z)||dz|=|E(w)||dw|
\end{equation}
\begin{equation}
|E(z)|^2|dA_z|=|E(w)|^2|dA_w|
\end{equation}
\begin{equation}
\label{eqn:Transform}
|E(z)|^2|dz| = |E(w)|^2|dw|\left|\frac{dw}{dz}\right| \, ,
\end{equation}
where the $w$-plane and the $z$-plane are related by conformal mapping, $|x|$ refers to taking the magnitude of $x$, $dw$ and $dz$ are small length elements in the $w$-plane and $z$-plane respectively, and $dA_w$ and $dA_z$ are small area elements in the $w$-plane and $z$-plane respectively.

\subsection{Multilayer Weighted-Average Approximation}
\label{sec:eff_con}
The superconductor pair-breaking potential $\Delta_g$, for a specific layer-thickness configuration, is sometimes obtained without solving the Usadel equations directly (for example via a look-up table/plot, or through experimental measurements), or is obtained by solving the Usadel equations only at frequencies close to the gap (to significantly reduce computation time). In these cases, the following approximation may be used to obtain the transmission line behaviour of superconducting multilayers. The effective conductivity is calculated using
\begin{align}
\label{eqn:approx}
\sigma_e=\frac{1}{t_{total}}\sum_{i}\sigma_i t_i \, ,
\end{align}
where $t_{total}=\sum_{i}t_i$, and suffix $i$ indicates the ith layer. This is the intuitive result which can be obtained from considering a parallel combination of layers. From $\sigma_e$, one can proceed with the surface impedance calculations \textit{as if} the material is homogeneous and has superconducting gap $\Delta_g$, normal state conductivity $\sigma_e$, and thickness $t_{total}$.

As we demonstrate, the homogeneous approximation fits well with the full calculation. The sub-gap behaviour of proximity-coupled transmission lines can now be effectively modelled without having to perform the full calculations outlined in section~\ref{sec:Impedances}.

\section{Microstrip Transmission Line Calculations}
\label{sec:Microstrip}
\subsection{Conformal Mapping}
\begin{figure}[!h]
\includegraphics[width=8.6cm]{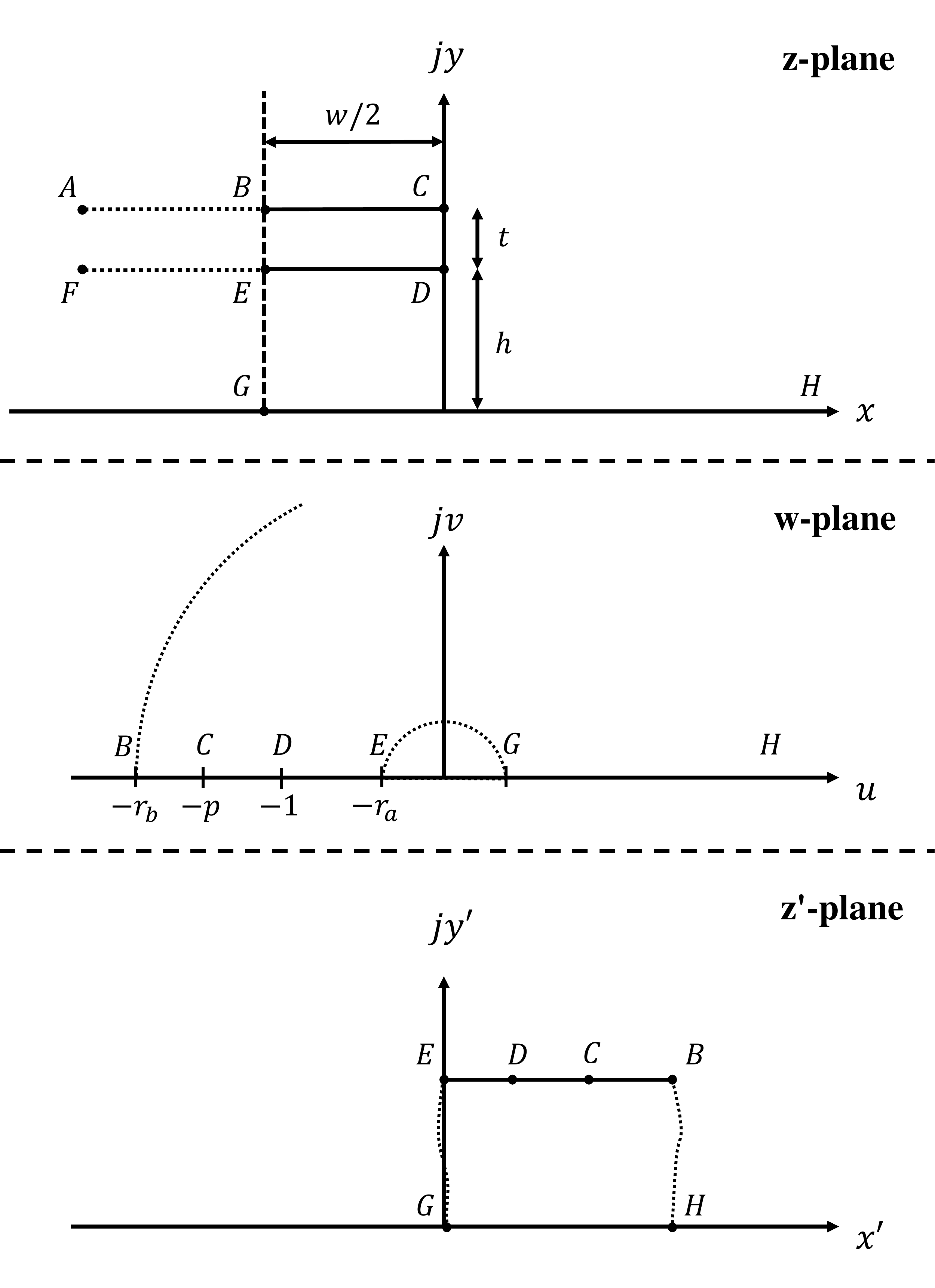}
\caption{\label{fig:conformalMicrostrip} Conformal mapping transformations for a microstrip transmission line. Here $z=x+jy$, $w=u+jv$, and $z'=x'+jy'$. Top: microstrip transmission line in $z$-plane; Mid: microstrip transmission line transformed into slot-line in the $w$-plane; Bottom: slot-line transformed into parallel plates in the $z'$-plane.}
\end{figure}
Two Schwarz-Christoffel transformations, detailed in \citenumns{Chang_1976,Chang_1979}, are needed to analyse a finite thickness MTL. With reference to figure~\ref{fig:conformalMicrostrip}, a MTL with width $w$, thickness $t$, and dielectric height $h$ in the $z$-plane is transformed into an infinitely thin, narrow gap slot-line in the $w$-plane. The metric coefficient of this transformation is
\begin{align}
\frac{dw}{dz}=\frac{\pi}{h}p^{1/2}\frac{w}{(w+1)^{1/2}(w+p)^{1/2}} \, ,
\end{align}
where $p= 2b^2-1+2b(b^2-1)^{1/2}$, and $b= 1+{t}/{h}$.

Afterwards, the slot-line is transformed into a pair of infinitely thin parallel plates in the $z'$-plane, with both plates having very similar plate width. The metric coefficient of this transformation is
\begin{align}
\frac{dz'}{dw}=\frac{1}{\pi}\frac{r_a(a^2-1)}{(r_a+aw)(ar_a+w)} \, ,
\end{align}
where $a=r_b/r_a+[{(r_b/r_a)^2-1}]^{1/2}$.

Here $r_a$ is closely approximated by
\begin{align}
\operatorname{ln}(r_a)= &-1-\frac{\pi w}{2h}-\frac{p+1}{p^{1/2}}\operatorname{tanh}^{-1}\left(p^{-1/2}\right) \\ \notag
&-\operatorname{ln}\left(\frac{p-1}{4p}\right) \, ,
\end{align}
and $r_b$ is closely approximated by
\begin{align}
r_b= r_{bo}
\end{align}
for $w/h\ge5$, and
\begin{align}
r_b= &r_{bo}-[(r_{bo}-1)(r_{bo}-p)]^{1/2} \\ \notag
&+(p+1)\operatorname{tanh}^{-1}\left(\frac{r_{bo}-p}{r_{bo}-1}\right)^{1/2} \\ \notag
&-2p^{1/2}\operatorname{tanh}^{-1}\left(\frac{r_{bo}-p}{p(r_{bo}-1)}\right)^{1/2}+\frac{\pi w}{2h}p^{1/2}
\end{align}
otherwise, where
\begin{align}
r_{bo}=&\Lambda+\frac{p+1}{2}\operatorname{ln}\Gamma \\
\Lambda=&p^{1/2}\left\{\frac{\pi w}{2h}+\frac{p+1}{2p^{1/2}}\left[1+\operatorname{ln}\left(\frac{4}{p-1}\right)\right]\right. \\ \notag
&\left.-2\operatorname{tanh}^{-1}p^{-1/2}\right\} \\
\Gamma=&\operatorname{max}(\Lambda,p) \, ,
\end{align}
and $\operatorname{max}(\Lambda,p)$ returns the larger of the two arguments.

\subsection{Evaluation of Geometric Factors}
The parallel-plate capacitance calculation has been performed by Chang \citenumns{Chang_1976,Chang_1979}, and is given by
\begin{equation}
C_{\text{vac}}=\frac{2\epsilon_0}{\pi}\operatorname{ln}\left(\frac{2r_b}{r_a}\right) .
\end{equation}
Using equation~(\ref{eqn:g1}), we thus have
\begin{align}
\label{eqn:MTL_g1}
g_1 = \frac{\pi}{2}\frac{1}{\operatorname{ln}\left({2r_b}/{r_a}\right)}
\end{align}

To find $\psi$, equation~(\ref{eqn:Psi}) is evaluated in the $w$-plane. We note here that $E(z')$ is approximately the electric field of a pair of non-fringing parallel plates, and that $r_b\gg r_a$. Hence the field in the $w$-plane is given by
\begin{align}
|E(w)|&=|E(z')|\left|\frac{dz'}{dw}\right| \\ \notag
&=\frac{hE_0}{\pi}\left|\frac{2r_b}{w(w+2r_b)}\right|  \, ,
\end{align}
where $E_0$ is the resultant constant field when the MTL width tends to infinity.
The denominator of $\psi$ can be directly calculated in the $z'$-plane, and is given by
\begin{align}
\int_{ext}|E|^2dA_{z'} = \frac{h^2E_0^2}{\pi}\operatorname{ln}\left(2\frac{r_b}{r_a}\right) .
\end{align}
We find the following contributions to $\psi_n$:
\begin{align}
\label{eqn:MTL_psi1}
\psi_u = &\frac{1}{2h}\frac{1}{\operatorname{ln}\left(2{r_b}/{r_a}\right)} \\ \notag
&\times\left[\int_{-p}^{-r_b}+\frac{1}{2}\int_{-1}^{-p}\right] \\ \notag
&\frac{4r_b^2p^{1/2}}{|(2r_b+w)^2w(w+1)^{1/2}(w+p)^{1/2}|}|dw| \\
\label{eqn:MTL_psi2}
\psi_l = &\frac{1}{2h}\frac{1}{\operatorname{ln}\left(2{r_b}/{r_a}\right)} \\ \notag
&\times\left[\int_{-r_a}^{-1}+\frac{1}{2}\int_{-1}^{-p}\right] \\ \notag
&\frac{4r_b^2p^{1/2}}{|(2r_b+w)^2w(w+1)^{1/2}(w+p)^{1/2}|}|dw| \\
\label{eqn:MTL_psi3}
\psi_g = &\frac{1}{2h}\frac{1}{\operatorname{ln}\left(2{r_b}/{r_a}\right)}\times \int_{r_a}^{r_b}\\ \notag
&\frac{4r_b^2p^{1/2}}{|(2r_b+w)^2w(w+1)^{1/2}(w+p)^{1/2}|}|dw| \, .
\end{align}
In accordance with \citenumns{Withington_1995}, we can approximate the $\psi_n$ integrals with the following analytic formulae.
\begin{align}
\psi_u &= \frac{I_u+\pi/2}{K_{MS}} \\
\psi_l &= \frac{I_l+\pi/2}{K_{MS}} \\
\psi_g &= \frac{I_g}{K_{MS}} \, ,
\end{align}
where
\begin{align}
I_u &= \operatorname{ln}\left(\frac{r_b(1-p)}{2p+2[p(1-r_b)(p-r_b)]^{1/2}-(p+1)r_b}\right) \\
I_l &=  \operatorname{ln}\left(\frac{2p+2[p(1-r_a)(p-r_a)]^{1/2}-(p+1)r_a}{r_a(p-1)}\right) \\
I_g &= \operatorname{ln}\left[\left(\frac{r_b}{2p+2[p(1+r_b)(p+r_b)]^{1/2}+(p+1)r_b}\right)\right. \notag \\
& \times \left.\left(\frac{2p+2[p(1+r_a)(p+r_a)]^{1/2}+(p+1)r_a}{r_a}\right)\right] \, ,
\end{align}
\begin{align}
K_{MS}= 2h\operatorname{ln}(r_b/r_a) \qquad &\text{for} \qquad w/h \,< \,2 \\ \notag
K_{MS}= 2h\operatorname{ln}(2r_b/r_a) \qquad &\textrm{otherwise}.
\end{align}
Using equation~(\ref{eq:Psi_defn}), $g_1$ and $\psi_{u,l,g}$ can then be substituted into equations~(\ref{eqn:Z},\ref{eqn:Y}) to compute the transmission line behaviour of the MTL.
\section{Coplanar Waveguide Calculations}
\label{sec:CPW}
\subsection{Conformal Mapping}
\begin{figure}[!h]
\includegraphics[width=8.6cm]{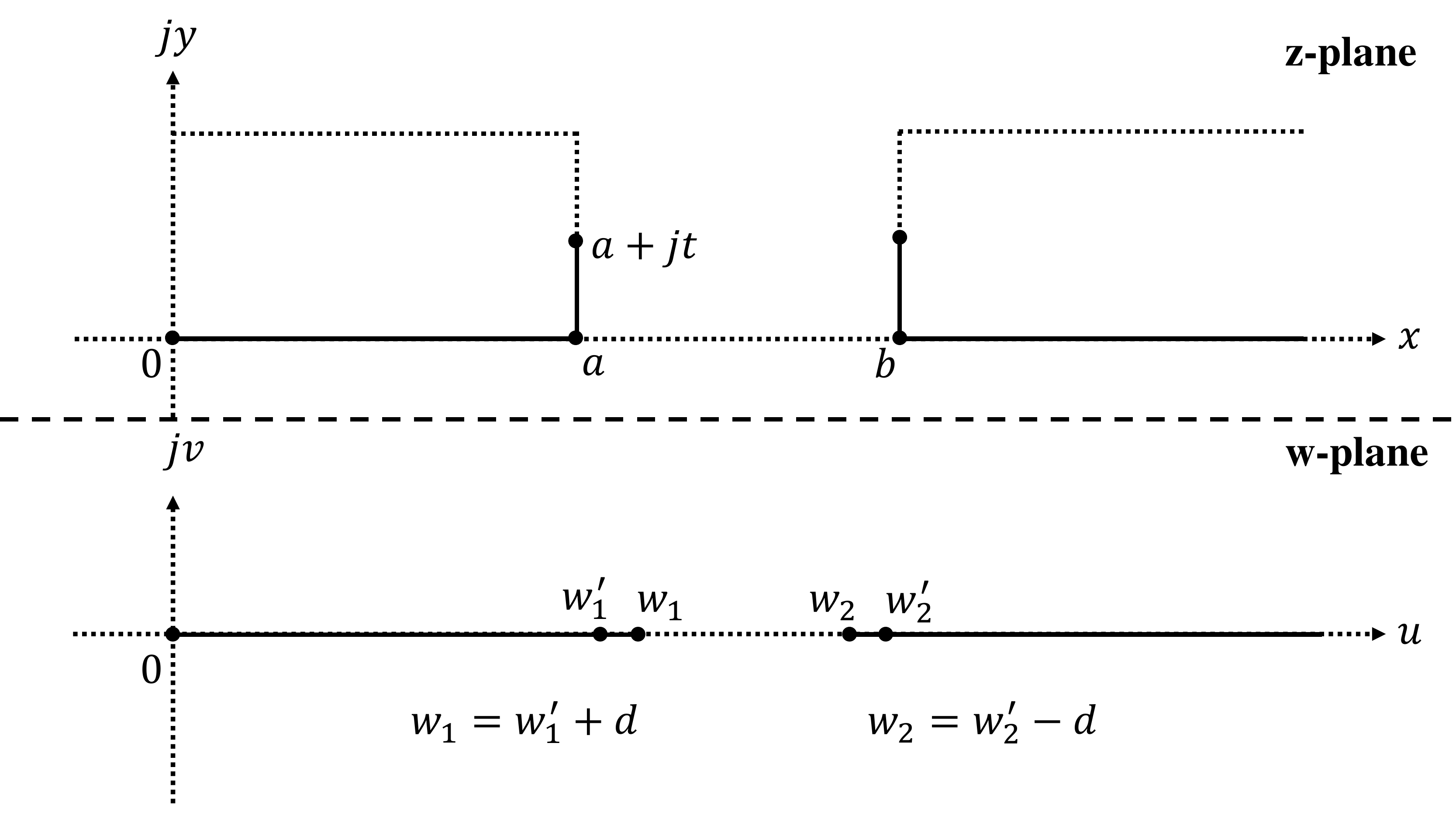}
\caption{\label{fig:conformalCPW1} The first conformal mapping for CPW which flattens the lower half of a finite thickness CPW into an infinitely thin CPW.  Here $z=x+jy$, $w=u+jv$, and $d$ is the length of the flattened CPW half-edge in the $w$-plane, and its value is given by equation~(\ref{eq:half_edge}).}
\end{figure}
Two Schwarz-Christoffel conformal transformations are needed to analyse a finite-thickness CPW. The first transformation, shown in figure~\ref{fig:conformalCPW1}, flattens the lower half of a CPW of inner strip width $2a$, ground plane separation $2b$, and thickness $2t$ into an infinitely thin CPW, using the transformation metric
\begin{equation}
\label{eqn:CPWtransformation}
\frac{dz}{dw}=\left[{\frac{(w^2-w_1'^2)(w^2-w_2'^2)}{(w^2-w_1^2)(w^2-w_2^2)}}\right]^{1/2} .
\end{equation}
To preserve symmetry, this transformation flattens half of the CPW along $y=t$.
\begin{figure}[!h]
\includegraphics[width=8.6cm]{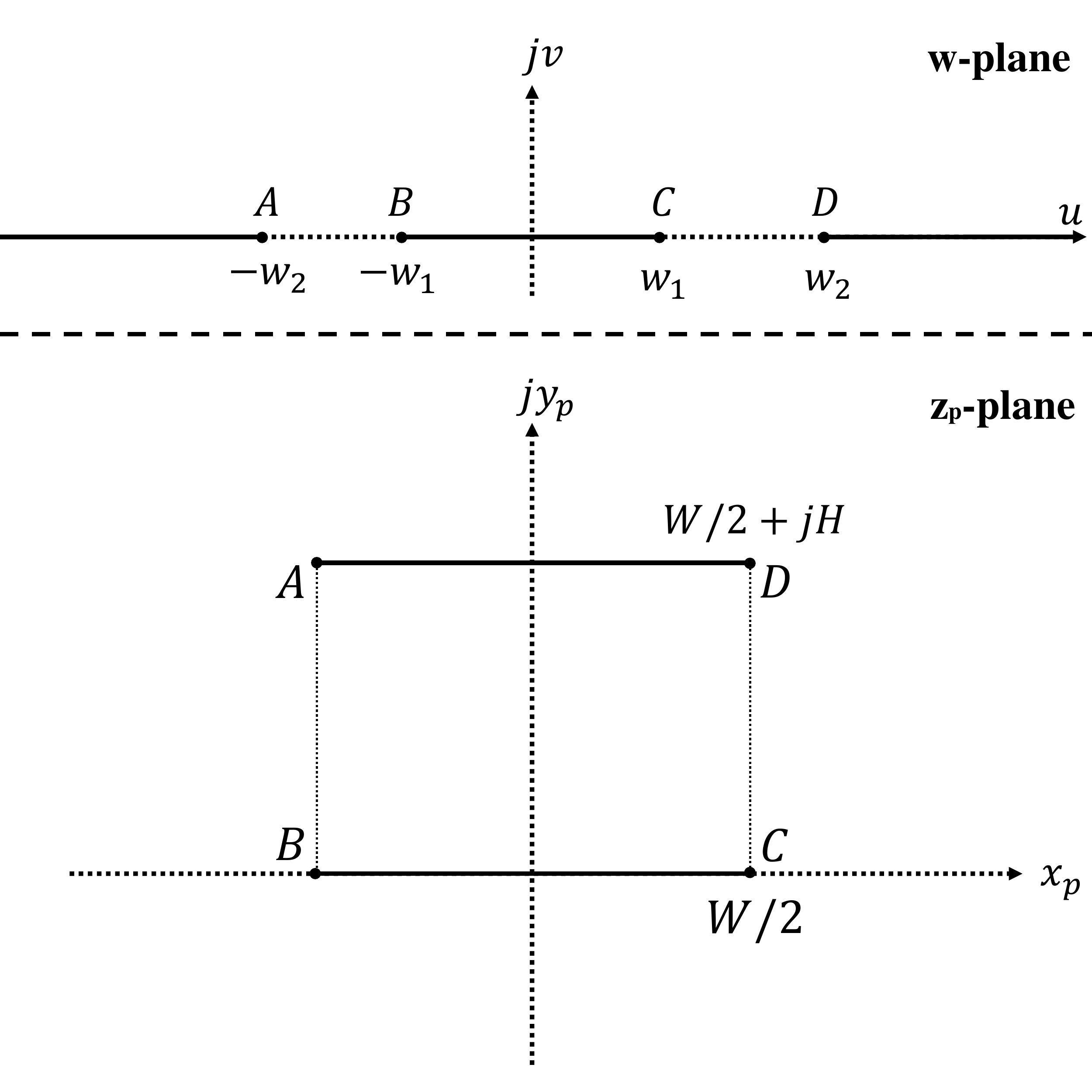}
\caption{\label{fig:conformalCPW2} The second conformal mapping for CPW which transforms an infinitely thin CPW into a pair of parallel plates.  Here $w=u+jv$ and $z_p=x_p+jy_p$. }
\end{figure}
The second transformation, shown in figure~\ref{fig:conformalCPW2}, transforms this thin CPW into a pair of parallel plates, using the transformation metric
\begin{equation}
\label{eqn:CPWtransformation2}
\frac{dz_p}{dw}=\left[{\frac{1}{(w^2-w_1^2)(w^2-w_2^2)}}\right]^{1/2} .
\end{equation}

The width and height of the parallel plates in the $z_p$-plane are given by $W= {2K(k)}/{b}$ and $H={K(k')}/{b}$ respectively, where $k=w_1/w_2$, $k'=({1-k^2})^{1/2}$, and $K(k)$ is the complete elliptical integral of the first kind.

In general, $w_1$, $w_1'$, $w_2$, and $w_2'$ have to be obtained by numerically solving the following system of equations, derived from equation~(\ref{eqn:CPWtransformation}):
\begin{align}
a &= \int_0^{w_1'} \frac{dz}{dw} dw \\
jt &= \int_{w_1'}^{w_1} \frac{dz}{dw} dw \\
b &= \int_{0}^{w_2'} \frac{dz}{dw} dw \\
jt &= \int_{w_2'}^{w_2} \frac{dz}{dw} dw .
\end{align}

In the case where the CPW thickness is much smaller than the gap width $b-a$, a good approximation of the points in the $w$-plane is given in \citenumns{Jiansong_2012}:
\begin{align}
w_1-w_1'&=w_2'-w_2=d=\frac{2}{\pi}t \label{eq:half_edge}\\
w_1 &= a+\frac{d}{2}-\frac{d}{2}\operatorname{ln}\frac{d}{a}+\frac{3}{2}d\operatorname{ln}2-\frac{d}{2}\operatorname{ln}\frac{a+b}{b-a} \\
w_2 &= b-\frac{d}{2}+\frac{d}{2}\operatorname{ln}\frac{d}{b}-\frac{3}{2}d\operatorname{ln}2+\frac{d}{2}\operatorname{ln}\frac{a+b}{b-a} .
\end{align}

\subsection{Evaluation of Geometric Factors}
The capacitance of an infinitely thin CPW in vacuo is well known \citenumns{Wen_1969}. Here we note that the \textit{pair} of CPW transformations should be applied \textit{twice} in total, for the upper \textit{and} the lower half of the finite thickness CPW. Hence the total capacitance is given by
\begin{align}
C_{\text{vac}} = 4\epsilon_0\frac{K(k)}{K(k')} .
\end{align}
An accurate approximation of the ratio ${K(k)}/{K(k')}$ can be found in \citenumns{Hilberg_1969}:
\begin{align}
\frac{K(k)}{K(k')}=
\frac{\pi}{\operatorname{ln}\left[2\frac{1+{k'}^{1/2}}{1-{k'}^{1/2}}\right]} \qquad &\text{for} \, 0\leq k\leq \frac{1}{\sqrt{2}} \\ \notag
\frac{K(k)}{K(k')}=\frac{\operatorname{ln}\left[2\frac{1+{k}^{1/2}}{1-{k}^{1/2}}\right]}{\pi} \qquad &\text{for} \,  \frac{1}{\sqrt{2}}\leq k\leq 1 .
\end{align}
Again using equation~(\ref{eqn:g1}), we have
\begin{align}
\label{eqn:CPW_g1}
g_1=\frac{K(k')}{4K(k)}
\end{align}

Before evaluating equation~(\ref{eqn:Psi}), it is important to note that each half plane of CPW contains a half of total power flow, a half of ground plane loss, and a single unit of upper \textit{or} lower strip loss.
The denominator of $\psi$ can be calculated in the $z_p$-plane, and is given by
\begin{align}
\int_{ext}|E|^2dA_{Z_p} &=\frac{2}{b^2} E_0^2 K(k)K(k') \, .
\end{align}
The numerator of equation~(\ref{eqn:Psi}) can then be evaluated in the $w$-plane using equation~(\ref{eqn:Transform}), which is the appropriate transformation rule. Numerically, we find the following contributions to $\psi$:
\begin{align}
\label{eqn:CPW_psi1}
\psi_u = \psi_l = \frac{b^2}{4K(k)K(k')}\int_0^{w_1}\left|\left(\frac{dz_p}{dw}\right)^2\frac{dw}{dz}\right|dw \\
\label{eqn:CPW_psi2}
\psi_g = \frac{b^2}{2K(k)K(k')}\int_{w_2}^{\infty}\left|\left(\frac{dz_p}{dw}\right)^2\frac{dw}{dz}\right|dw .
\end{align}
The above integrals can be approximated by a scheme detailed in \citenumns{Collin_2001, Owyang_1958}. We obtain
\begin{align}
\psi_{u,l} =& \frac{1}{8K(k)K(k')(1-k^2)}\notag \\
&\times\left[\frac{\pi}{a}+\frac{1}{a}\operatorname{ln}\left(\frac{8a}{d}\right)+\frac{1}{b}\operatorname{ln}\left(\frac{b-a}{b+a}\right)\right] \\
\psi_g = & \frac{1}{4K(k)K(k')(1-k^2)}\notag \\
&\times\left[\frac{\pi}{b}+\frac{1}{b}\operatorname{ln}\left(\frac{8b}{d}\right)+\frac{1}{a}\operatorname{ln}\left(\frac{b-a}{b+a}\right)\right] .
\end{align}
Using equation~(\ref{eq:Psi_defn}), $g_1$ and $\psi_{u,l,g}$ can then be substituted into equations~(\ref{eqn:Z},\ref{eqn:Y}) to compute the transmission line behaviour of the CPW.
\section{Computational Results}
\label{sec:Results}
The analysis described in the previous sections is applied to bilayer Al-Ti systems of variable thickness combinations. Al-Ti multilayers are chosen for this analysis because of their stability to corrosion, their long quasiparticle lifetimes, and their potential to be incorporated in important applications including: sub-$100\,\,\textrm{GHz}$ Cosmic Microwave Background observations \citenumns{Catalano_2015,Songyuan_2018}, low red-shift CO lines measurements at around $100-110\,\,\mathrm{GHz}$ \citenumns{Cicone_CO_2012,Thomas_2014,Songyuan_2018}, and phonon-mediated bolometric experiments searching for neutrinoless double-$\beta$ decay \citenumns{Cardani_2018}. Full numerical integration is performed here using MatLab without applying the analytic approximations.
\subsection{Physical Parameters}
\begin{figure}[!h]
\includegraphics[width=8.6cm]{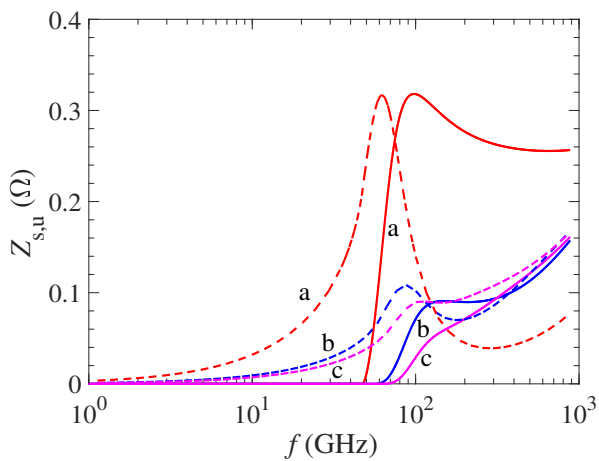}
\caption{\label{fig:Zs} Surface impedance $Z_{s,u}$ of Al-Ti bilayers with $t_{Ti}=100\,\textrm{nm}$, $T=0.1\,\,{\textrm K}$ and $\gamma_{\mathrm{B,Al}}=0.01$. The Al layer is the upper layer. Solid lines indicate surface resistance $R_s$; dashed lines indicate surface reactance $X_s$. (a) red line,~$t_{\mathrm{Al}}=25\,\textrm{nm}$, (b) blue line,~$t_{\mathrm{Al}}=100\,\textrm{nm}$, (c) magenta line,~$t_{\mathrm{Al}}=200\,\textrm{nm}$. $Z_{s,l}$ is qualitatively similar and is not shown for clarity.}
\end{figure}

\begin{table}[ht]
\begin{threeparttable}
\caption{\label{tab:table1}Table of material properties.}
\begin{tabular}{b{0.50\linewidth} b{0.26\linewidth} b{0.18\linewidth}}
\toprule
 & \textrm{Aluminium} & \textrm{Titanium} \\
\colrule
$T_{\mathrm {c}}$ (K) & 1.2\tnote{a} & 0.55\tnote{a} \\
$\sigma_{\mathrm{N}}$ (/$\mu\Omega\,$m) \tnote{b} & 132\tnote{a} & 5.88\tnote{a} \\
$RRR$ \tnote{c} & 5.5\tnote{a} &  3.5\tnote{a} \\
$n_0$ ($10^{47}$/$\textrm{J}\,\textrm{m}^3$) & 1.45\tnote{d} & 1.56\tnote{d} \\
$D$ ($\mathrm{m^2s^{-1}}$) & 35\tnote{e} &  1.5\tnote{e} \\
$\xi$ (nm) & 189\tnote{f} & 57\tnote{f} \\
\toprule
\end{tabular}
\begin{tablenotes}[flushleft]
\RaggedRight
\footnotesize
\item[a] Measurements by the Cambridge Quantum Sensors Group. \\
\item[b] $\sigma_{\mathrm{N}}$ is the normal state conductivity. \\
\item[c] $RRR$ is the residual resistivity ratio.\\
\item[d] $n_0$ is the normal state electron density of states, and is calculated from the free electron model \citenumns{Ashcroft_1976}. \\
\item[e] Diffusivity constant $D$ is calculated using $D_s = \sigma_{N,s}/(n_{0,s}e^2)$ \citenumns{Martinis_2000}. \\
\item[f] Coherence length $\xi$ is calculated using $\xi_s=[{\hbar D_s/(2\pi k_\mathrm {B} T_\mathrm {c}})]^{1/2}$ \citenumns{Brammertz2001}, where $k_\mathrm {B}$ is the Boltzmann constant.
\end{tablenotes}
\end{threeparttable}
\end{table}
Table \ref{tab:table1} presents the physical parameters used in numerical calculations. Using the method discussed in \citenumns{Songyuan_2018}, we calculate $Z_{s,u}$ and $Z_{s,l}$ for bilayer configurations with $t_{Ti}=100\,\textrm{nm}$, $t_{Al}=25, 100, 200\,\textrm{nm}$, where $t_{Ti}$ indicates thickness of the titanium layer, and $t_{Al}$ indicates thickness of the aluminum layer. Typical behaviour of $Z_{s,u}$ is shown in figure~\ref{fig:Zs}. Using these results, we carry out the transmission line calculations discussed in previous sections. Calculations are performed at $T=0.1\,\,{\textrm K}$, $\gamma_{\mathrm{B,Al}}=0.01$, where $\gamma_{\mathrm{B,Al}}=R_{\mathrm{B}}\sigma_{N,Al}/\xi_{Al}$, $R_{\mathrm{B}}$ is the product of the boundary resistance between the Al-Ti layers and the boundary area. $\gamma_{\mathrm{B,Al}}=0.01$ is chosen as a representative value of general clean boundary parameters.

\subsection{Bilayer Microstrip Transmission Line Results}
The MTL geometry explored here has dimensions $w=2 \,\mu\textrm{m}$, $h=300 \,\textrm{nm}$, $\epsilon_r=3.8$ representative of $\mathrm{SiO}_2$, and ground plane thickness $t_{g}=1\,\mu\textrm{m}$. Total MTL thickness is given by the sum of individual layer thicknesses $t_{total}=t_{Al}+t_{Ti}$. $\epsilon_{fm}(f)$ is taken from Edwards' and Owens' empirical formula \citenumns{Edwards_1976}. The ground plane material is aluminium.

\begin{figure}[!h]
\includegraphics[width=8.6cm]{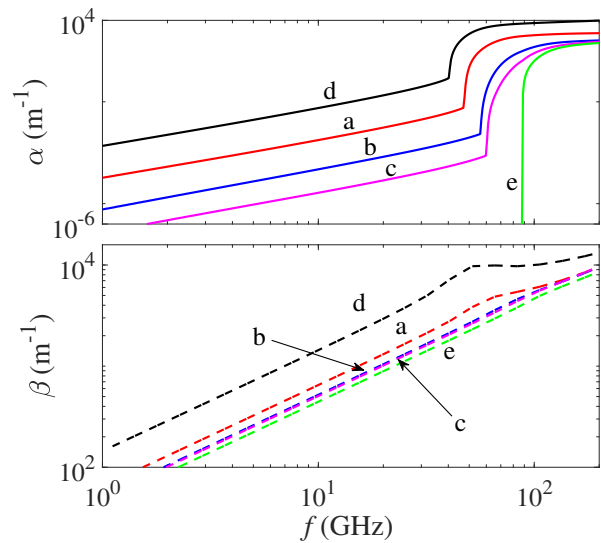}
\caption{\label{fig:MS_alphabeta} Propagation constant $\gamma=\alpha+j\beta$ of Al-Ti bilayer microstrip transmission lines against frequency $f$ for $t_{Ti}=100\,\textrm{nm}$, $T=0.1\,\,{\textrm K}$ and $\gamma_{\mathrm{B,Al}}=0.01$. The Al layer is the upper layer. Solid lines in the upper figure indicate $\alpha$; dashed lines in the lower figure indicate $\beta$. (a) red line,~$t_{\mathrm{Al}}=25\,\textrm{nm}$, (b) blue line,~$t_{\mathrm{Al}}=100\,\textrm{nm}$, (c) magenta line,~$t_{\mathrm{Al}}=200\,\textrm{nm}$, (d) black line, pure titanium at $100\,\textrm{nm}$, (e) green line, pure aluminium at $10\,\mu\textrm{m}$. }
\end{figure}

Our bilayer system has two homogeneous asymptotic limits according to the thickness combinations explored here: 100 nm pure Ti and infinitely thick pure Al. Figure~\ref{fig:MS_alphabeta} confirms that the bilayer propagation constants are indeed in between their homogeneous asymptotic values. The attenuation constants $\alpha$ (solid lines in the upper plot) are strongly frequency dependent, sharply increasing in magnitude above the pair-breaking frequencies.

%\begin{figure}[!h]
%\includegraphics[width=8.6cm]{02_lossTan.pdf}
%\caption{\label{fig:MS_lossTan} Loss tangent $\operatorname{tan}\delta$ against frequency $f$ for $t_{Ti}=100\,\textrm{nm}$, $T=0.1\,\,{\textrm K}$ and $\gamma_{\mathrm{B,Al}}=0.01$, where Al layer is the upper layer. (a) red line,~$t_{\mathrm{Al}}=25\,\textrm{nm}$, (b) blue line,~$t_{\mathrm{Al}}=100\,\textrm{nm}$, (c) magenta line,~$t_{\mathrm{Al}}=200\,\textrm{nm}$, (d) black line, pure titanium at $100\,\textrm{nm}$, (e) green line, pure aluminium at $10\,\mu\textrm{m}$. }
%\end{figure}

\begin{figure}[!h]
\includegraphics[width=8.6cm]{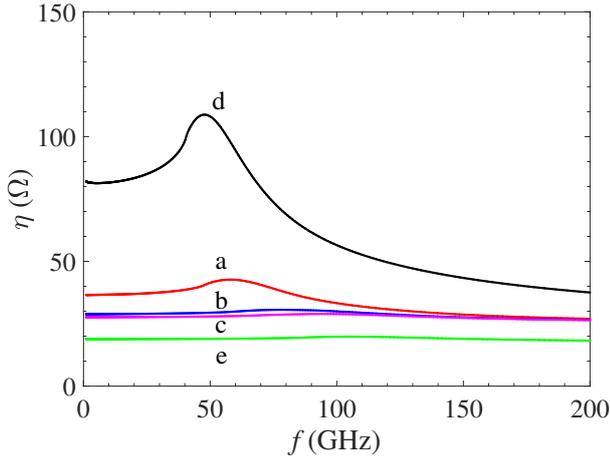}
\caption{\label{fig:MS_eta} Characteristic impedance $\eta$ of Al-Ti bilayer microstrip transmission lines against frequency $f$ for $t_{Ti}=100\,\textrm{nm}$, $T=0.1\,\,{\textrm K}$ and $\gamma_{\mathrm{B,Al}}=0.01$. The Al layer is the upper layer. (a) red line,~$t_{\mathrm{Al}}=25\,\textrm{nm}$, (b) blue line,~$t_{\mathrm{Al}}=100\,\textrm{nm}$, (c) magenta line,~$t_{\mathrm{Al}}=200\,\textrm{nm}$, (d) black line, pure titanium at $100\,\textrm{nm}$, (e) green line, pure aluminium at $10\,\mu\textrm{m}$. }
\end{figure}

Figure~\ref{fig:MS_eta} shows the calculated characteristic impedances of the bilayer MTLs studied. The bilayer impedances lie in between their BCS asymptotes. Such calculations allow the possibility of tuning a superconducting transmission line using proximity layers to achieve the desired characteristic impedances (for example to $\mathrm{50\,\Omega}$). Line (d) demonstrates the strongest frequency-dependent features. This is because its series impedance is dominated by the strongly frequency-dependent surface impedance term, i.e. the second term of equation~(\ref{eqn:Z}). In contrast, line (e) is almost featureless. This is because its series impedance is dominated by the frequency-independent geometry term, i.e. the first term of equation~(\ref{eqn:Z}).

%\begin{figure}[!h]
%\includegraphics[width=8.6cm]{06_eta_approx.pdf}
%\caption{\label{fig:MS_eta_approx_f} Characteristic impedance $\eta$ against frequency $f$ for $t_{Ti}=100\,\textrm{nm}$, $t_{Al}=100\,\textrm{nm}$, $T=0.1\,\,{\textrm K}$ and $\gamma_{\mathrm{B,Al}}=0.01$. (a) red line, Al layer on top, (b) black line, Ti layer on top, (c) blue line, effective conductivity approximation.}
%\end{figure}

\begin{figure}[!h]
\includegraphics[width=8.6cm]{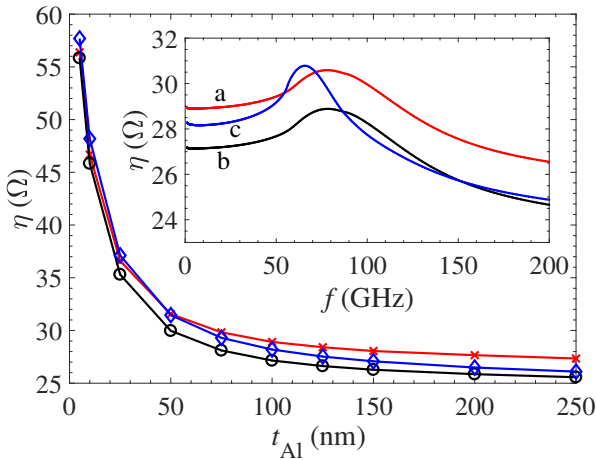}
\caption{\label{fig:MS_eta_approx_3GHz} Characteristic impedance $\eta$ of Al-Ti bilayer microstrip transmission lines against Al thickness $t_{Al}$ for $t_{Ti}=100\,\textrm{nm}$, $T=0.1\,\,{\textrm K}$, $\gamma_{\mathrm{B,Al}}=0.01$, at $f =3\,\textrm{GHz}$. (a) cross, red line, Al layer on top, (b) circle, black line, Ti layer on top, (c) diamond, blue line, effective conductivity approximation. Inset: $\eta$ against frequency $f$ for $t_{Ti}=100\,\textrm{nm}$, $t_{Al}=100\,\textrm{nm}$ device. (a) red line, Al layer on top, (b) black line, Ti layer on top, (c) blue line, effective conductivity approximation. }
\end{figure}

Figure~\ref{fig:MS_eta_approx_3GHz} demonstrates the effective conductivity approximation described in section~\ref{sec:eff_con}. The main plot focuses on the sub-gap behaviour of approximated characteristic impedances, plotted against different thickness configurations. $3\,\,{\mathrm {GHz}}$ is used here as a representative value of sub-gap frequencies relevant to KIDs and qubit operations. The approximation works well throughout the range of Al thicknesses. The approximation is most accurate when $t_{Al}\approx t_{Ti}$. The inset shows that the effective conductivity approximation works well across a wide range of frequencies. A homogeneous material has no preferred orientation. Hence the \textit{best} approximation should lie in between the full calculations with opposite layer ordering. Slight deviation occurs at frequencies close to the pair-breaking frequencies. This is due to the fact that proximity-coupled multilayers have significantly different densities of states (DoS) close to the pair-breaking energy gap. A more detailed discussion on the shape of DoS in multilayers can be found in \citenumns{Songyuan_2018}.

\begin{figure}[!h]
\includegraphics[width=8.6cm]{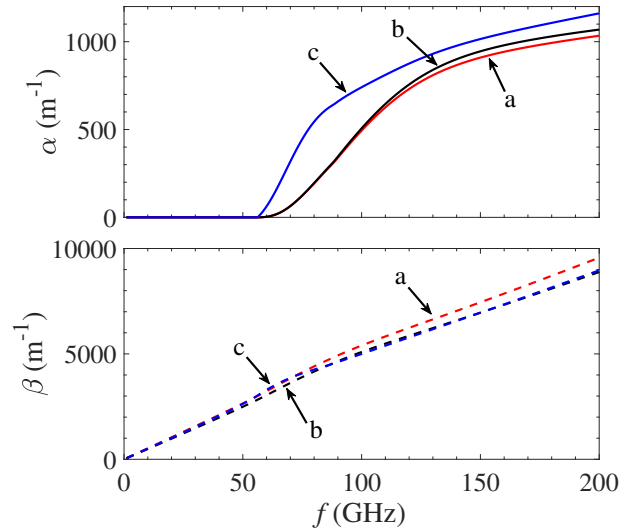}
\caption{\label{fig:MS_a_b_approx} Propagation constant $\gamma=\alpha+j\beta$ of Al-Ti bilayer microstrip transmission lines against frequency $f$ for $t_{Ti}=100\,\textrm{nm}$, $t_{Al}=100\,\textrm{nm}$, $T=0.1\,\,{\textrm K}$ and $\gamma_{\mathrm{B,Al}}=0.01$. The Al layer is the upper layer. Solid lines in the upper figure indicate $\alpha$; dashed lines in the lower figure indicate $\beta$. (a) red line, Al layer on top, (b) black line, Ti layer on top, (c) blue line, effective conductivity approximation. }
\end{figure}

Figure~\ref{fig:MS_a_b_approx} shows the propagation constants obtained from the effective conductivity approximation, and that obtained from full calculations of multilayer surface impedances. $\beta$ is well approximated, whilst $\alpha$ demonstrates a similar degree of deviation compared to the approximations of $\eta$ shown in figure~\ref{fig:MS_eta_approx_3GHz}. Since similar deviation information is conveyed by plots of $\eta$ and plots of the propagation constants, the figures in later discussions will focus mainly on the behaviour of $\eta$.

\begin{figure}[!h]
\includegraphics[width=8.6cm]{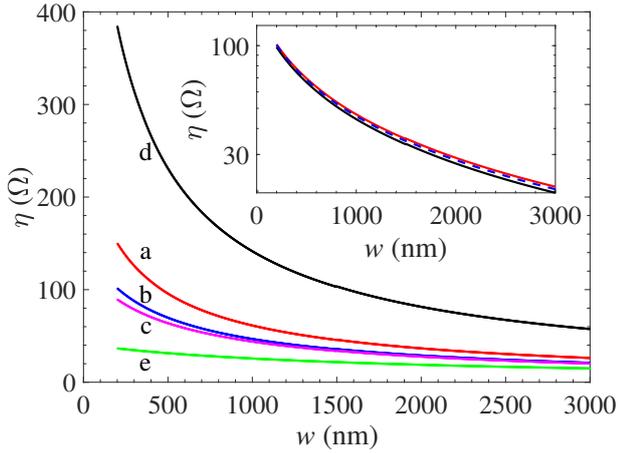}
\caption{\label{fig:MS_eta_w} Characteristic impedance $\eta$ of Al-Ti bilayer microstrip transmission lines against conductor strip width $w$ for $t_{Ti}=100\,\textrm{nm}$, $T=0.1\,\,{\textrm K}$, $\gamma_{\mathrm{B,Al}}=0.01$, at $f =3\,\textrm{GHz}$. The Al layer is the upper layer. (a) red line,~$t_{\mathrm{Al}}=25\,\textrm{nm}$, (b) blue line,~$t_{\mathrm{Al}}=100\,\textrm{nm}$, (c) magenta line,~$t_{\mathrm{Al}}=200\,\textrm{nm}$, (d) black line, pure titanium at $100\,\textrm{nm}$, (e) green line, pure aluminium at $10\,\mu\textrm{m}$. Inset: $\eta$ against conductor strip width $w$ for $t_{Ti}=100\,\textrm{nm}$, $t_{\mathrm{Al}}=100\,\textrm{nm}$, $T=0.1\,\,{\textrm K}$, $\gamma_{\mathrm{B,Al}}=0.01$, at $f =3\,\textrm{GHz}$. (a) solid red line, Al layer on top, (b) solid black line, Ti layer on top, (c) dashed blue line, effective conductivity approximation.}
\end{figure}

Figure~\ref{fig:MS_eta_w} shows the width dependence of the characteristic impedance $\eta$, at $f =3\,\textrm{GHz}$. The inset shows that the effective conductivity approximation works well throughout a wide range of MTL widths. The 3 GHz impedances remain in between the values obtained from reversing the layer ordering.

%\begin{figure}[!h]
%\includegraphics[width=8.6cm]{09_w_eta_approx.pdf}
%\caption{\label{fig:MS_eta_w_approx} Characteristic impedance $\eta$ against conductor strip width $w$ for $t_{Ti}=100\,\textrm{nm}$, $t_{\mathrm{Al}}=100\,\textrm{nm}$, $T=0.1\,\,{\textrm K}$, $\gamma_{\mathrm{B,Al}}=0.01$, at $f =3\,\textrm{GHz}$. (a) red line, Al layer on top, (b) black line, Ti layer on top, (c) blue line, effective conductivity approximation.}
%\end{figure}

\subsection{Bilayer Coplanar Waveguide Results}
The CPW geometry explored here has dimensions $a=1 \,\mu\textrm{m}$, $b=2 \,\mu\textrm{m}$, $\epsilon_r=11.7$ representative of $\textrm{Si}$, and substrate height $h=200\,\mu\textrm{m}$. The CPW central strip thickness is given by the sum of the individual layer thicknesses $t_{total}=t_{Al}+t_{Ti}=2t$. The ground strip has the same thickness as the conductor strip. $\epsilon_{fm}(f)$ is computed from a tabulation of formulae in chapter 7 of \citenumns{Gupta_1996}, taking into account finite substrate thicknesses, non-zero strip thicknesses, and dispersive behaviour. Note that due to the symmetry of the conformal mapping \textit{in vacuo}, the transmission line properties of a CPW bilayer are not affected by layer orientation. The behaviour of $\gamma$ is qualitatively similar to that of MTL, and is not shown here.

%\begin{figure}[!h]
%\includegraphics[width=8.6cm]{11_eta.pdf}
%\caption{\label{fig:CPW_eta} Characteristic impedance $\eta$ against frequency $f$ for $t_{Ti}=100\,\textrm{nm}$, $T=0.1\,\,{\textrm K}$ and $\gamma_{\mathrm{B,Al}}=0.01$. (a) red line,~$t_{\mathrm{Al}}=25\,\textrm{nm}$, (b) blue line,~$t_{\mathrm{Al}}=100\,\textrm{nm}$, (c) magenta line,~$t_{\mathrm{Al}}=200\,\textrm{nm}$, (d) black line, pure titanium at $100\,\textrm{nm}$, (e) green line, pure aluminium at $500\,\textrm{nm}$. }
%\end{figure}

\begin{figure}[!h]
\includegraphics[width=8.6cm]{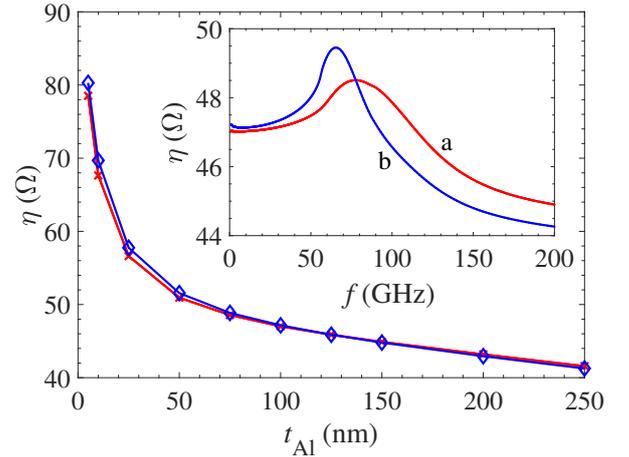}
\caption{\label{fig:CPW_eta_approx} Characteristic impedance $\eta$ of Al-Ti bilayer coplanar waveguides against Al thickness $t_{Al}$ for $t_{Ti}=100\,\textrm{nm}$, $T=0.1\,\,{\textrm K}$, $\gamma_{\mathrm{B,Al}}=0.01$, at $f =3\,\textrm{GHz}$. (a) cross, red line, full calculation, (b) diamond, blue line, effective conductivity approximation. Inset: $\eta$ against frequency $f$ for $t_{Ti}=100\,\textrm{nm}$, $t_{Al}=100\,\textrm{nm}$. (a) red line, full calculation, (b) blue line, effective conductivity approximation.}
\end{figure}

The effective conductivity approximation is applied to CPW geometry, and compared against the full calculation in figure~\ref{fig:CPW_eta_approx}. The approximation performs better for CPW than for MTL. This is primarily due to the fact that CPW geometry is not sensitive to layer orientation. The homogeneous approximation is intrinsically orientation-insensitive, and thus performs better when approximating an orientation-insensitive geometry as opposed to an orientation-sensitive geometry.

\begin{figure}[!h]
\includegraphics[width=8.6cm]{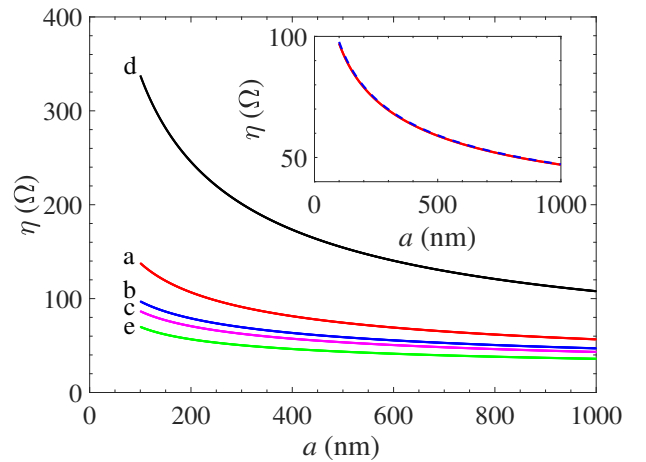}
\caption{\label{fig:CPW_eta_a} Characteristic impedance $\eta$ of Al-Ti bilayer coplanar waveguides against conductor strip half-width $a$ for $t_{Ti}=100\,\textrm{nm}$, $T=0.1\,\,{\textrm K}$, $\gamma_{\mathrm{B,Al}}=0.01$, at $f =3\,\textrm{GHz}$. (a) red line,~$t_{\mathrm{Al}}=25\,\textrm{nm}$, (b) blue line,~$t_{\mathrm{Al}}=100\,\textrm{nm}$, (c) magenta line,~$t_{\mathrm{Al}}=200\,\textrm{nm}$, (d) black line, pure titanium at $100\,\textrm{nm}$, (e) green line, pure aluminium at $10\,\mu\textrm{m}$. Inset: $\eta$ against conductor strip half-width $a$ for $t_{Ti}=100\,\textrm{nm}$, $t_{\mathrm{Al}}=100\,\textrm{nm}$, $T=0.1\,\,{\textrm K}$, $\gamma_{\mathrm{B,Al}}=0.01$, at $f =3\,\textrm{GHz}$. (a) solid red line, full calculation, (b) dashed blue line, effective conductivity approximation. }
\end{figure}

In the next set of calculations, the width of the conductor strip has been varied across the range of $2a=200-2000 \,\textrm{nm}$, whilst the gap has been kept constant at $b-a=1 \,\mu\textrm{m}$. At comparable total conductor strip widths ($w$ for MTL and $2a$ for CPW), CPW has consistently higher characteristic impedances, as seen in comparing figure~\ref{fig:CPW_eta_a} with figure~\ref{fig:MS_eta_w}. Further, the inset shows that the effective conductivity approximation works well across a wide range of strip widths.

%\begin{figure}[!h]
%\includegraphics[width=8.6cm]{15_a_eta_approx.pdf}
%\caption{\label{fig:CPW_eta_a_approx} Characteristic impedance $\eta$ against half conductor strip width $a$ for $t_{Ti}=100\,\textrm{nm}$, $t_{\mathrm{Al}}=100\,\textrm{nm}$, $T=0.1\,\,{\textrm K}$, $\gamma_{\mathrm{B,Al}}=0.01$, at $f =3\,\textrm{GHz}$. (a) black line, full calculation, (b) blue line, effective conductivity approximation.}
%\end{figure}

\section{Discussion and Conclusions}
\label{sec:Conclusions}
In this study, we have presented a treatment of superconducting MTL and CPW that is suitable for homogeneous superconductors as well as proximity-coupled multilayers. Our analysis is based on conformal mapping and takes into account the thicknesses of transmission lines. We have also presented a simple weighted-average approximation, which allows a multilayer to be modelled by a homogeneous material with an effective conductivity.
Our results for MTL converge to that of \citenumns{Withington_1995} in the limit that the conductor strip and the ground plane are made of the same homogeneous material. Our analytic forms of CPW attenuation constants agree with the results in \citenumns{Owyang_1958,Collin_2001} for a homogeneous CPW completely surrounded by a single dielectric material, the results in \citenumns{Clem_2013} when $b-a \ll a$, and the results in \citenumns{Holloway_1995} for rectangular CPW edges. These agreements in the appropriate limits give confidence to the validity of this work. In addition, our treatment of CPW gives the full propagation constants which are not found in previous works.

In summary, series impedances and shunt admittances are calculated from equations~(\ref{eqn:Z}-\ref{eqn:Y}). The effective conductivity approximation is given in equation~(\ref{eqn:approx}). For MTL devices, $g_1$ is calculated from equation~(\ref{eqn:MTL_g1}), and $\psi_n$ is calculated from equations~(\ref{eqn:MTL_psi1}-\ref{eqn:MTL_psi3}). For CPW devices, $g_1$ is calculated from equation~(\ref{eqn:CPW_g1}), and $\psi_n$ is calculated from equations~(\ref{eqn:CPW_psi1}-\ref{eqn:CPW_psi2}).

We have applied our analysis routine to Al-Ti bilayers of varying thicknesses, and found that our effective conductivity approximation works well for both MTL and CPW geometries.

Future design procedures of superconducting MTL/CPW can utilize this analysis routine to reduce the time and resources spent on empirical characterisations. Calculations of characteristic impedances could be used in computer-aided design packages, and could benefit systems that require impedance matching; calculations of loss tangents allow predictions on transmission line sensitivity; calculations of series impedances and shunt admittances allow modelling of transmission line systems via telegrapher's equations, for example, in parametric amplifier simulations \citenumns{Shan_2016}.

The analysis framework of this study, in the context of homogeneous MTL, has previously been applied and verified by various experiments \citenumns{Visser_2014,Yassin_2000,Shan_2007}. We thus expect this work to be similarly applicable and valuable to future scientific studies involving superconducting transmission lines.

\bibliographystyle{IOP_no_URL}
\bibliography{library}

\providecommand{\newblock}{}
\begin{thebibliography}{10}
\expandafter\ifx\csname url\endcsname\relax
  \def\url#1{{\tt #1}}\fi
\expandafter\ifx\csname urlprefix\endcsname\relax\def\urlprefix{URL }\fi
\providecommand{\eprint}[2][]{\url{#2}}

\bibitem{Jonas_review}
Zmuidzinas J {2012} {\em {Annu. Rev. Condens. Matter Phys.}\/} {\bf {3}}
  169--214

\bibitem{Johansson_2006}
Johansson G, Tornberg L and Wilson C~M 2006 {\em Phys. Rev. B\/} {\bf 74}(10)
  100504

\bibitem{Majer_2007}
Majer J, Chow J~M, Gambetta J~M, Koch J, Johnson B~R, Schreier J~A, Frunzio L,
  Schuster D~I, Houck A~A, Wallraff A, Blais A, Devoret M~H, Girvin S~M and
  Schoelkopf R~J 2007 {\em Nature\/} {\bf 449} 443 EP --

\bibitem{Wesenberg_2009}
Wesenberg J~H, Ardavan A, Briggs G~A~D, Morton J~J~L, Schoelkopf R~J, Schuster
  D~I and M\o{}lmer K 2009 {\em Phys. Rev. Lett.\/} {\bf 103}(7) 070502

\bibitem{Pla_2016}
Pla J~J, Bienfait A, Pica G, Mansir J, Mohiyaddin F~A, Zeng Z, Niquet Y~M,
  Morello A, Schenkel T, Morton J~J~L and Bertet P 2016 Strain-induced spin
  resonance shifts in silicon devices (\textit{Preprint}
  \eprint{arXiv:1608.07346})

\bibitem{Bienfait_2015}
Bienfait A, Pla J~J, Kubo Y, Stern M, Zhou X, Lo C~C, Weis C~D, Schenkel T,
  Thewalt M~L~W, Vion D, Esteve D, Julsgaard B, M{\o}lmer K, Morton J~J~L and
  Bertet P 2015 {\em Nat. Nanotechnol.\/} {\bf 11} 253 EP --

\bibitem{Eom_2012}
Eom B~H, Day P~K, LeDuc H~G and Zmuidzinas J 2012 {\em Nat. Phys.\/} {\bf 8}
  623

\bibitem{Trucker_1985}
Tucker J~R and Feldman M~J 1985 {\em Rev. Mod. Phys.\/} {\bf 57}(4) 1055--1113

\bibitem{WEDENIG_1999}
Wedenig R, Niinikoski T, Berglund P, Kyynäräinen J, Costa L, Valtonen M,
  Linna R, Salmi J, Seppä H and Suni I 1999 {\em Nucl. Instrum. Meth. A\/}
  {\bf 433} 646 -- 663

\bibitem{Doyle_2008}
Doyle S 2008 {\em Lumped Element Kinetic Inductance Detectors\/} Ph.D. thesis
  Cardiff University

\bibitem{CALDER_2016}
Vignati M, Bellini F, Cardani L, Casali N, Castellano M~G, Colantoni I,
  Coppolecchia A, Cosmelli C, Cruciani A, D’Addabbo A, Domizio S~D, Martinez
  M and Tomei C 2016 {\em J. Phys. Conf. Ser.\/} {\bf 718} 062065

\bibitem{Moore_2012}
Moore D~C 2012 {\em A search for low-mass dark matter with the cryogenic dark
  matter search and the development of highly multiplexed phonon-mediated
  particle detectors\/} Ph.D. thesis California Institute of Technology

\bibitem{Catalano_2015}
Catalano A, Goupy J, le~Sueur H, Benoit A, Bourrion O, Calvo M, D'addabbo A,
  Dumoulin L, Levy-Bertrand F, Macias-Perez J, Marnieros S, Ponthieu N and
  Monfardini A {2015} {\em {Astron. Astrophys.}\/} {\bf {580}} {A15}

\bibitem{Songyuan_2018}
Zhao S, Goldie D~J, Withington S and Thomas C~N 2018 {\em Supercond. Sci.
  Tech.\/} {\bf 31} 015007

\bibitem{Cardani_2018}
Cardani L, Casali N, Cruciani A, le~Sueur H, Martinez M, Bellini F, Calvo M,
  Castellano M~G, Colantoni I, Cosmelli C, D'Addabbo A, Domizio S~D, Goupy J,
  Minutolo L, Monfardini A and Vignati M 2018 Al/ti/al phonon-mediated kids for
  uv-vis light detection (\textit{Preprint} \eprint{arXiv:1801.08403})

\bibitem{Kaplan_1979}
Kaplan S~B 1979 {\em J. Low Temp. Phys.\/} {\bf 37} 343--365

\bibitem{Thorbeck_2015}
Thorbeck T and Zimmerman N~M 2015 {\em AIP Advances\/} {\bf 5} 087107

\bibitem{Withington_1995}
Yassin G and Withington S 1995 {\em J. Phys. D Appl. Phys.\/} {\bf 28} 1983

\bibitem{Visser_2014}
de~Visser P~J, Goldie D~J, Diener P, Withington S, Baselmans J~J~A and Klapwijk
  T~M 2014 {\em Phys. Rev. Lett.\/} {\bf 112}(4) 047004

\bibitem{Yassin_2000}
Yassin G, Withington S, Buffey M, Jacobs K and Wulff S 2000 {\em IEEE T.
  Microw. Theory Techn.\/} {\bf 48} 662--669

\bibitem{Shan_2007}
Shan W, Shi S, Matsunaga T, Takizawa M, Endo A, Noguchi T and Uzawa Y 2007 {\em
  IEEE T. Appl. Supercon.\/} {\bf 17} 363--366

\bibitem{Mattis_1958}
Mattis D~C and Bardeen J 1958 {\em Phys. Rev.\/} {\bf 111}(2) 412--417

\bibitem{Kautz_1978}
Kautz R~L 1978 {\em J. Appl. Phys.\/} {\bf 49} 308--314

\bibitem{Usadel_1970}
Usadel K~D 1970 {\em Phys. Rev. Lett.\/} {\bf 25} 507--509

\bibitem{Nam_1967}
Nam S~B {1967} {\em {Phys. Rev.}\/}

\bibitem{Whitaker_1988}
Whitaker J~F, Sobolewski R, Dykaar D~R, Hsiang T~Y and Mourou G~A 1988 {\em
  IEEE T. Microw. Theory Techn.\/} {\bf 36} 277--285

\bibitem{Edwards_1976}
Edwards T~C and Owens R~P 1976 {\em IEEE T. Microw. Theory Techn.\/} {\bf 24}
  506--513

\bibitem{Gupta_1996}
Gupta K 1996 {\em Microstrip Lines and Slotlines\/} Artech House Atennas And
  Propagation Library (Artech House)

\bibitem{Chang_1979}
Chang W~H 1979 {\em J. Appl. Phys.\/} {\bf 50} 8129--8134

\bibitem{Assadourian_1952}
Assadourian F and Rimai E 1952 {\em Proc. IRE\/} {\bf 40} 1651--1657

\bibitem{Chang_1976}
Chang W~H 1976 {\em IEEE T. Microw. Theory Techn.\/} {\bf 24} 608--611

\bibitem{Jiansong_2012}
Gao J 2008 {\em The physics of superconducting microwave resonators\/} Ph.D.
  thesis California Institute of Technology

\bibitem{Wen_1969}
Wen C~P 1969 {\em IEEE T. Microw. Theory Techn.\/} {\bf 17} 1087--1090

\bibitem{Hilberg_1969}
Hilberg W 1969 {\em IEEE T. Microw. Theory Techn.\/} {\bf 17} 259--265

\bibitem{Collin_2001}
Collin R 2001 {\em Foundations for Microwave Engineering\/} IEEE Press Series
  on Electromagnetic Wave Theory (Wiley)

\bibitem{Owyang_1958}
Owyang G and Wu T 1958 {\em IRE Trans. Anntenas Propag.\/} {\bf 6} 49--55

\bibitem{Cicone_CO_2012}
Cicone C, Feruglio C, Maiolino R, Fiore F, Piconcelli E, Menci N, Aussel H and
  Sturm E {2012} {\em {Astron. Astrophys.}\/} {\bf {543}} {A99}

\bibitem{Thomas_2014}
Thomas C~N, Withington S, Maiolino R, Goldie D~J, {de Lera Acedo} E, Wagg J,
  Blundell R, Paine S and Zeng L {2014}  (\textit{Preprint}
  \eprint{{astro-ph.IM/1401.4395v1}})

\bibitem{Ashcroft_1976}
Ashcroft N and Mermin N 1976 {\em {Solid State Physics}\/} (Philadelphia:
  Saunders College)

\bibitem{Martinis_2000}
Martinis J~M, Hilton G, Irwin K and Wollman D 2000 {\em Nucl. Instrum. Meth.
  A\/} {\bf 444} 23 -- 27

\bibitem{Brammertz2001}
Brammertz G, Golubov A, Peacock A, Verhoeve P, Goldie D~J and Venn R 2001 {\em
  Physica C Supercond.\/} {\bf 350} 227--236

\bibitem{Clem_2013}
Clem J~R 2013 {\em J. Appl. Phys.\/} {\bf 113} 013910 (\textit{Preprint}
  \eprint{https://doi.org/10.1063/1.4773070})

\bibitem{Holloway_1995}
Holloway C~L and Kuester E~F 1995 {\em IEEE T. Microw. Theory Techn.\/} {\bf
  43} 2695--2701

\bibitem{Shan_2016}
Shan W, Sekimoto Y and Noguchi T 2016 {\em IEEE T. Appl. Supercon.\/} {\bf 26}
  1--9

\end{thebibliography}
\end{document}